# Wave and ray analysis of a type of cloak exhibiting magnified and shifted scattering effect

Yu Luo [1], Jingjing Zhang [1], Hongsheng Chen [1,3]*, Bae-Ian Wu[3], Lixin Ran [1,2], and Jin Au Kong [1,3]

[1] *The Electromagnetics Academy at Zhejiang University, Zhejiang University, Hangzhou 310027, China*

[2] *Department of Information and Electronics Engineering, Zhejiang University, Hangzhou 310027, China*

[3] *Research Laboratory of Electronics, Massachusetts Institute of Technology, Cambridge, Massachusetts 02139, USA*

**Abstract**

Ray-tracing exercise and full-wave analysis were performed to validate the performance of a new type of cloak composed of isotropic metamaterials. It is shown that objects inside the 'folded region' of this cloak appear invisible to the incoming light from a ray tracing exercise, but exhibit magnified and shifted scattering under a plane wave illumination from a full wave analysis. Gaussian beams are introduced to resolve this interesting paradox resulted from these two methods. We show that at the time-harmonic state, small energy can be diffracted into the folded region and contribute to the resonant state even when the Gaussian beam is steered away from the cloak with an object inside. A scattering pattern identical to that scattered from the image of the object will be formed, which agrees well with the phenomenon in the plane wave incidence case.

*Corresponding author, Email: chenhs@ewt.mit.edu

Cloaking an obstacle from electromagnetic wave (EM) detection has attracted much interest. To make an object invisible, we need to guarantee the scattering cross section of the object as small as possible. Alù and Engheta presented the technique of using plasmonic and metamaterial covers to reduce the total scattering of a single or a collection of subwavelength spheres [1, 2], which relies on an inherently nonresonant scattering cancellation phenomenon; In Ref. [3], a method of concealing polarizable lines or point dipoles situated near a cloak made of a superlens was suggested by Milton *et. al.* [3]; Based on the form invariance of Maxwell's equations, Pendry *et*. *al*. proposed the optical transformation approach to design invisibility cloak with zero scattering cross section [4]. This methodology has recently won much more concerns [5-15], since wave cannot penetrate into the core of the cloaking device and the cloak can effectively render object of any shape and size invisible.

Different from previous efforts which try to conceal an object by minimizing its scattering, we propose an inverse way to camouflage the object by magnifying and shifting the scattering from the obstacle so that it seems to be produced by a much larger scatterer deviated from its original position. The proposed cloak is based on the super-imagining effect of left-handed medium (LHM) [16] coating and can be used to disguise the true nature of the object, e.g. the position, the size, etc, and further mislead the detector who uses the time-harmonic EM waves. However, ray approximation method predicts a different behavior, showing that light can be bent around the 'folded region' of the cloak, making the object inside invisible. Gaussian beam which has a flexibility to be modeled as a ray (i.e. narrow Gaussian beam) or a plane wave (i.e. wide Gaussian beam) is implemented to resolve this paradox. We show that at time-harmonic state, even if the narrow Gaussian beam (or ray) is shifted from the cloak, small energy can still be diffracted into the folded region, inducing the resonance inside. Therefore when an object is located inside the coating layer, a scattering pattern identical to that scattered from the image of the object will be formed under a time-harmonic wave incidence. Apart from the anomalous phenomenon brought about by this cloak, it is also worth mentioning that this type of cloak can be implemented by isotropic negative index materials with the magnitude of the spatially varying refractive index larger than one everywhere. In addition, the material parameters of the cloak do not have extreme value (zero or infinity) anywhere. Therefore, concealment can be achieved with artificially structured negative-index media (NIM) like phonic crystal [17] or subwavelength dielectric resonator [18].

For simplification, we suppose the spherical cloaking shell is a shell with inner radius $R_1$ and outer radius $R_2$. Under a radial mapping $r' = f(r)$, the material parameters of the three-dimensional (3D) transformation medium will take the form

$$\varepsilon_r = \mu_r = \frac{f^2(r)}{r^2 f'(r)}, \quad \varepsilon_\theta = \mu_\theta = \varepsilon_\varphi = \mu_\varphi = f'(r). \tag{1}$$

The invisibility property requires $f(R_1) = 0$. Under this condition, it can be easily concluded that the parameters are strongly anisotropic and the radial components are zero at the inner surface. In order to reduce the anisotropy of the cloak, we choose $f(r)$ to be $f_1(r) = R_2^2 / r$, which is monotonically decreasing as depicted in Fig. 1(a) ($f(r) = r$ represents the free space in the outer region). The corresponding permittivity and permeability become $\varepsilon_1 = \mu_1 = -R_2^2 / r^2$, which means the coating is constructed by spatially variant isotropic LHM [16] and the magnitude of the parameters are all larger than 1 (since $R_1 < r < R_2$). In order to make the impedance of the inner core matched with that of the coating layer, we let the material in the inner core be a RHM with the parameters: $\varepsilon_2 = \mu_2 = (R_2 / R_1)^2$. We can consider the LHM coating along with the RHM core as a whole concealing device, whose parameters can be achieved by a non-monotonic transformation function shown in Fig. 1 (a). It should be noted that in the folded region $R_0 < r < R_2$, a certain value of $f(r)$ corresponds to two different radii. In fact, many interesting phenomena are correlated with that region, which will be addressed later.

The function of the cloak is first demonstrated by using a ray tracing exercise. By taking the geometry limit of Maxwell's equations, the light path of the ray can be obtained with numerical integration of a set of Hamiltion's equations [4, 5]. The results are shown in Fig. 1(b). The light is incident from the bottom and propagating along $z$ direction. For comparison, we have also included the ray tracing results when the light is incident onto Pendry's cloak, which is shown in Fig. 1(c). Different from Pendry's perfect cloak which guides the rays around the central volume, our cloak in Fig. 1(b) guides the light to penetrate the core, but bends it all around the aforementioned 'folded region'. Therefore, an object located in the folded region of the cloak should be invisible to ray detection from the ray tracing analysis.

The Ray tracing exercise in the above section shows that the cloak achieved by a non-monotonic transformation function shown in Fig.1 (a) can smoothly guide the incoming light around the 'folded region', making an obstacle here effectively invisible. However, totally different phenomena can be observed under a full wave analysis. Suppose a linearly polarized plane wave with unit amplitude $E^{\text{inc}} = \hat{x}e^{ik_0 z}$ is incident upon the coating along the $z$ direction. As explained by Fig. 2, a perfect electric conducting (PEC) sphere with radius $R_s$ is located at the position $\bar{r}_0(r_0, \theta_0, \varphi_0)$ in the folded region ($R_0 < r < R_2$) of the coating. Here, $\bar{r}_0$ is a position vector whose three coordinate parameters are $r_0$, $\theta_0$, $\varphi_0$, respectively. We use Mie scattering theory to study this case [19]. Similar to the process in Ref. [10] and [14], by describing the fields with two Bromwich scalar potentials $\Phi_{TE}$ and $\Phi_{TM}$ and applying the boundary conditions, the scattered potentials can be obtained as:

$$\Phi_{TE/TM}^{s} = E_0 \sum_{n=1}^{\infty}\sum_{m=-n}^{n}\sum_{l=1}^{\infty} a_n T_n^{TE/TM} C_{n,m,l}^{TE/TM} \psi_n\left(k_0 r_0 \frac{R_2^2}{R_1^2}\right) \zeta_n(k_0 r) P_n^m(\cos\xi) e^{im(\varphi-\varphi_0)}, \qquad (2)$$

where $\cos\xi = \cos\theta\cos\theta_0 + \sin\theta\sin\theta_0\cos(\varphi-\varphi_0)$, and $a_n = \dfrac{(-i)^{-n}(2n+1)}{n(n+1)}$.

$T_n^{TM} = -a_n \dfrac{\psi_n'\left[k_0 R_s (R_2/R_1)^2\right]}{\zeta_n'\left[k_0 R_s (R_2/R_1)^2\right]}$ and $T_n^{TE} = -a_n \dfrac{\psi_n\left[k_0 R_s (R_2/R_1)^2\right]}{\zeta_n\left[k_0 R_s (R_2/R_1)^2\right]}$ are the scattering coefficients. And $C_{n,m,l}^{TE/TM}$ is a definite expansion coefficient [19]. $\psi_n$ and $\zeta_n$ represent the Riccati-Bessel function of the first and the third kind, respectively. $P_n^m$ is the $n$-th orders of the associated Legendre polynomials of degree $m$. It is worth mentioning that Eq. (2) is nonconvergent in the annulus region $R_2 < r < (R_2/R_1)^2 r_0$ due to the negative index of the material in the region $R_1 < r < R_2$, similar to the perfect lens imaging case in principle [20, 21]. In the domain $r > (R_2/R_1)^2 r_0$, Eq. (2) can be written in the following closed forms:

$$\begin{cases} \Phi_{TM}^{s} = \dfrac{\cos\varphi'}{\omega} \sum_{n=1}^{\infty} a_n T_n^{TM} \zeta_n\left(k_0 \left|\overline{r} - (R_2/R_1)^2 \overline{r_0}\right|\right) P_n^1(\cos\theta') \\ \Phi_{TE}^{s} = \dfrac{\sin\varphi'}{\omega\eta_0} \sum_{n=1}^{\infty} a_n T_n^{TE} \zeta_n\left(k_0 \left|\overline{r} - (R_2/R_1)^2 \overline{r_0}\right|\right) P_n^1(\cos\theta') \end{cases}, \qquad (3)$$

where $\overline{r}'(r',\theta',\varphi') = \overline{r} - (R_2/R_1)^2 \overline{r_0}$ is a definite position vector (shown in Fig. 2). Eq. (3) shows that the scalar potentials of the whole system is exactly the same as that of a PEC sphere with radius $R_s (R_2/R_1)^2$ located at a position $(R_2/R_1)^2 \overline{r_0}$. Fig. 3 (a) depicts the total field distribution of such case obtained by full wave analysis, where the frequency of the incident plane wave is 2 GHz. The radii of the inner and outer boundaries of the coating are 5 *cm* and 10 *cm* respectively, and a PEC sphere with the radius $R_s$ =1.5 *cm* is located at point A (corresponding to point A in Fig. 1(a)) inside the cloak. A magnified image is formed at point B (corresponding to point B in Fig.1 (a)). Fig. 3(b) displays the case where a conductive sphere with radius 6 *cm* is placed at point **B** in free space. By comparison, we find that the field distributions in the two cases are identical, indicating that an observers outside will 'see' a bigger scatterer (with a size magnification of 4 in this case) which has been shifted from its real position. In this way, both the volume and position information of the object will be disguised, misleading the observer effectively. This interesting phenomenon can be understood by checking Fig. 1(a) again. Note that at point A and point B, *f(r)* have the same value, indicating that an object at point A of the cloaking device will cause the identical scattering to that from a larger object (with a size magnification of $\left(R_2/R_1\right)^2$) located at point **B** in free space (*r* > $R_2$). One thing worth pointing out is that, it takes some time for the field distribution shown in Fig. 3(a) to be established. In fact, under the outer illumination, the coating will concentrate the incoming wave into the internal region before reaching the steady state. During this period, the whole system will store energy from the incident wave, making some energy circulate in the folded region. Thanks to this energy circulation, the scattering shift and enhancement phenomena are gradually formed in the time harmonic state. When the obstacle inside the cloak is absent, as shown in Fig. 3 (c), the energy is still circulating in the folded region without introducing any scattering.

The results of the ray tracing method and the full wave analysis method are seemingly inconsistent. The paradox comes from the fact that the ray tracing method has not considered the diffraction of the light into the folded region. In order to further understand this phenomenon from physical perspective, we consider the case where a

Gaussian beam is incident upon the cloak in the free space. The fields of the incident beam can be decomposed into TE and TM modes as

$$\Phi^i_{\text{TE/TM}} = E_0 \sum_{n=1}^{\infty} \sum_{m=-n}^{n} D^{\text{TE/TM}}_{n,p,w} g^{\text{TE/TM}}_{n,m} \psi_n(k_0 r) P_n^m(\cos\theta) e^{im\varphi} \quad (4)$$

where $D^{\text{TE/TM}}_{n,p,w}$ and $g^{\text{TE/TM}}_{n,m}$ are the expansion coefficients which can be determined through the approach of Ref. [22]. The scattered field can be deduced by solving the boundary equations. Suppose the inner and outer radius of the coating is $R_1 = 20\lambda$ and $R_2 = 40\lambda$, and the PEC obstacle with radius $R = 2\lambda$ is located at the folded region. Note that here the wavelength is very small compared with the size of the cloak. Instead of achieving scattering shift and enhancement, the whole cloaking device compresses the incoming Gaussian beam and guide it around the folded region, as depicted in Fig. 4(a). In other words, the scattered field outside the cloak approaches to zero and the obstacle in the folded region is invisible to outer observers. This phenomenon of wave bending is similar to what we find in the ray tracing method, where all the waves incident upon the outer surface of the LHM coating have been concentrated in a small region ($r < R_0$, $R_0$ is shown in Fig. 1(a)) as they penetrate into the core. In plane wave incidence case, the power of the incident wave will diffract into the folded region due to the high spatial dispersion. And the energy will be stored there and circulate between the LHM coating and the RHM core. This is also the reason why the power flows inside the core is larger than the power flows through the whole system [14]. However, in the cases where the spatial dispersion of the incident beam is so small that we can neglect the power dispersed into the cloak, the detecting beam will transmit through the cloak without touching the obstacle in the folded region. In order to observe the effect of the dispersed power more clearly, we remove the scatterer and move the Guassian beam away from the cloak, and then calculate the field distribution inside and outside the cloak. Fig. 4(b) shows the field distribution obtained by analytical calculation. We can observe that in the steady state, a resonance is induced by the dispersed field and the power is circulated in the folded region. In fact, since the field dispersed into the cloak is very weak, it takes quite a long time for this steady state to be established. Next, we replace the beam in Fig. 4(a) with a wide Gaussian beam, as shown in Fig. 4 (c). It is found that, when the width of the Gaussian beam is comparable with the size of the cloak, very similar field distribution to that of the plane wave case (Fig. 3(a)) is formed. Similarly, we also change the Gaussian beam in Fig. 4 (b) for a wide beam, which is shown in Fig. 4(d). The same as in Fig. 3(c), the wave is concentrated in the center of the cloak.

The above analysis help us better understand the seemingly contradiction between the ray tracing and full-wave methods. The narrow Gaussian beam analysis shows high agreement with the ray tracing exercise in most of the cases, where the beam incident upon the cloak will all be bent around the folded region without causing scattering. However, different from what we find from the ray tracing method, a narrow Gaussian beam passing by the cloak will cause an inherent resonance of the cloak as long as the time is long enough. Therefore, the resonances of the ray in the negative refractive index media need to be included in the ray tracing model in order to solve this paradox. Similar contradiction of the ray and wave analysis can also be shown in a configuration of negative space composed of negative refractive index materials [16]. In our case we use the narrow Gaussian beam, which can be vividly viewed as a ray, as shown in Fig. 4(b), and we can find the internal resonance of the ray in the interface of the positive and negative index materials.

Another point worth noting is that the whole cloak is constructed by isotropic media. In geometric optics limit, the effective refraction index is the main factor we consider which can affect the light paths. It is required that the refraction index of the LH coating and the RH core are $n_1 = -R_2^2/r^2$ and $n_2 = (R_2/R_1)^2$ respectively (these parameters are also applicable to two dimensional cylindrical case), which is physically realizable with metamaterials, such as photonic crystal [17] and high dielectric resonators [18]. The reflection at the interfaces can be suppressed by controlling the thickness and surface termination of the LHM coating layer.

In summary, analytical full wave theory and ray tracing method have been introduced to examine the behavior of a new type of cloak consisting of a spherical LHM coating and an RHM core. It is demonstrated that this cloak will lead to a scattering enhancement instead of scattering reduction in contrast to all the traditional cloaks; the scattered field distribution can also be changed as if the scatterer is moved away from its original position. Thanks to those capabilities, the proposed cloak can be used to disguise the true information of the object, and further mislead the observer. Our approach provides a new perspective to achieve camouflage, and is an alternative to the published invisibility cloaking methods.

**Acknowledgement**

This work is sponsored by the National Science Foundation of China under grants 60801005 and 60531020, in part by the Zhejiang National Science Foundation under grant R1080320 and Y1080715, the Ph.D Programs Foundation of MEC (No. 20070335120 and 200803351025), the NCET-07-0750, 863 Project (No. 2009AA01Z227), the Office of Naval Research (ONR) under contract N00014-06-1-0001, and the Department of the Air Force under Air Force contract FA8721-05-C-0002.

**Figure captions**

Figure 1. (color online) (a) Schematic figure of the functions $f(r)$ which is non-monotonic. The inner media ($0 < r < R_1$) is RHM, while the media in the coating ($R_1 < r < R_2$) is LHM. The fold ($R_0 < r < R_2$) indicates the hidden region. (b) Ray trajectories when the scatterer is located at the hidden region of the coating. The light has been smoothly bent around this region. (c) The path of the rays when a scatterer is placed in an ideal transformation cloak proposed by Pendry. No light can enter into the center of the shell.

Figure 2. (color online) (a) 3D geometry of an obstacle (red solid object) with a radius $R_s = 1.5$cm located inside the cloak at the position $\vec{r}_0(r_0, \theta_0, \varphi_0)$. The transparent red object with dashed contour stands for the image which will be formed under an EM wave illumination.

Figure 3. (color online) (a) Field distribution of an $E_x$ polarized plane wave incident upon the coating along $z$ direction when a conductive sphere with radius 1.5 cm is placed at the position **A** in the hidden region of coating. (b) Field distribution of the same plane wave incident upon a conductive sphere with radius 6 cm placed at the position **B** in free space. (c) Field distribution of the plane wave incident upon the coating when there are no scatterers inside the coating.

Figure 4. (color online) (a) Snapshot of the electric field distribution when a narrow Gaussian beam is incident upon the cloak with $R_2 = 40\lambda$ and $R_1 = 20\lambda$ along the $z$ direction. A PEC scatterer is located inside the folded region of the cloak (b) The electric field distribution when a narrow Gaussian beam is shifted away from the cloak. A resonance can be observed inside the folded region, which is due to the accumulation of the dispersed field from the Gaussian beam. (c) Similar to case (a), but incident with a wide Gaussian beam. (d) Similar to case (b), but incident with a wide Gaussian beam.

Figure 1

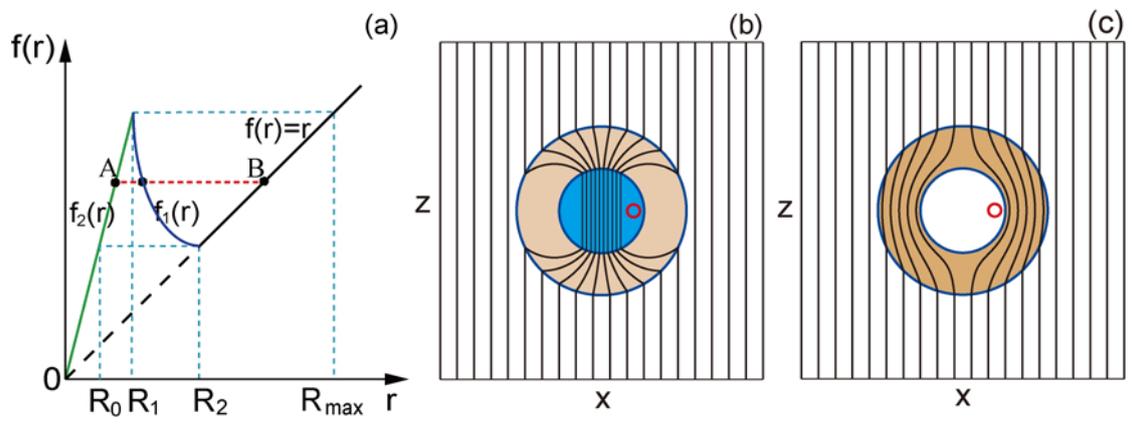

Figure 2

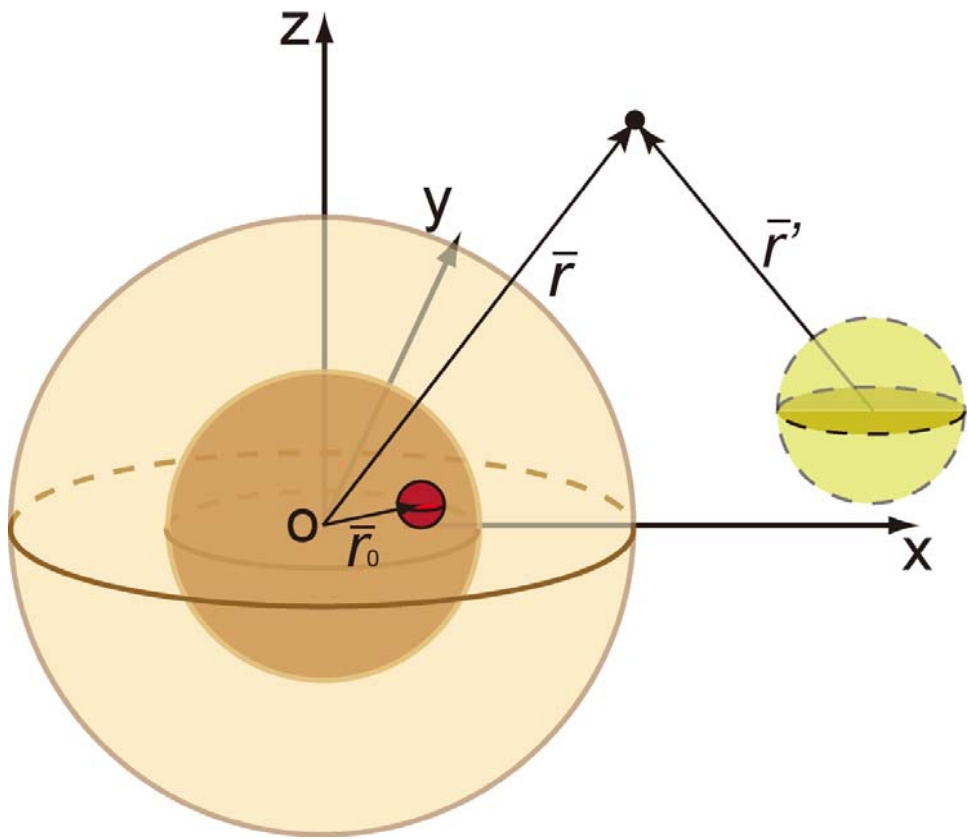

Figure 3

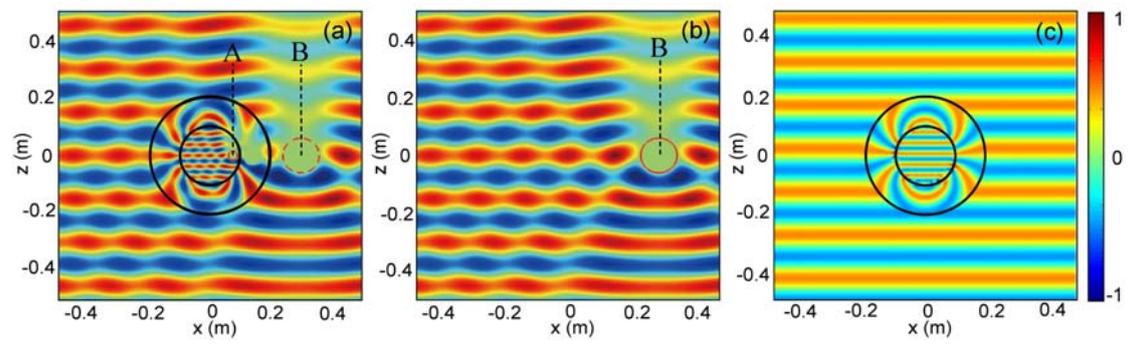

Figure 4

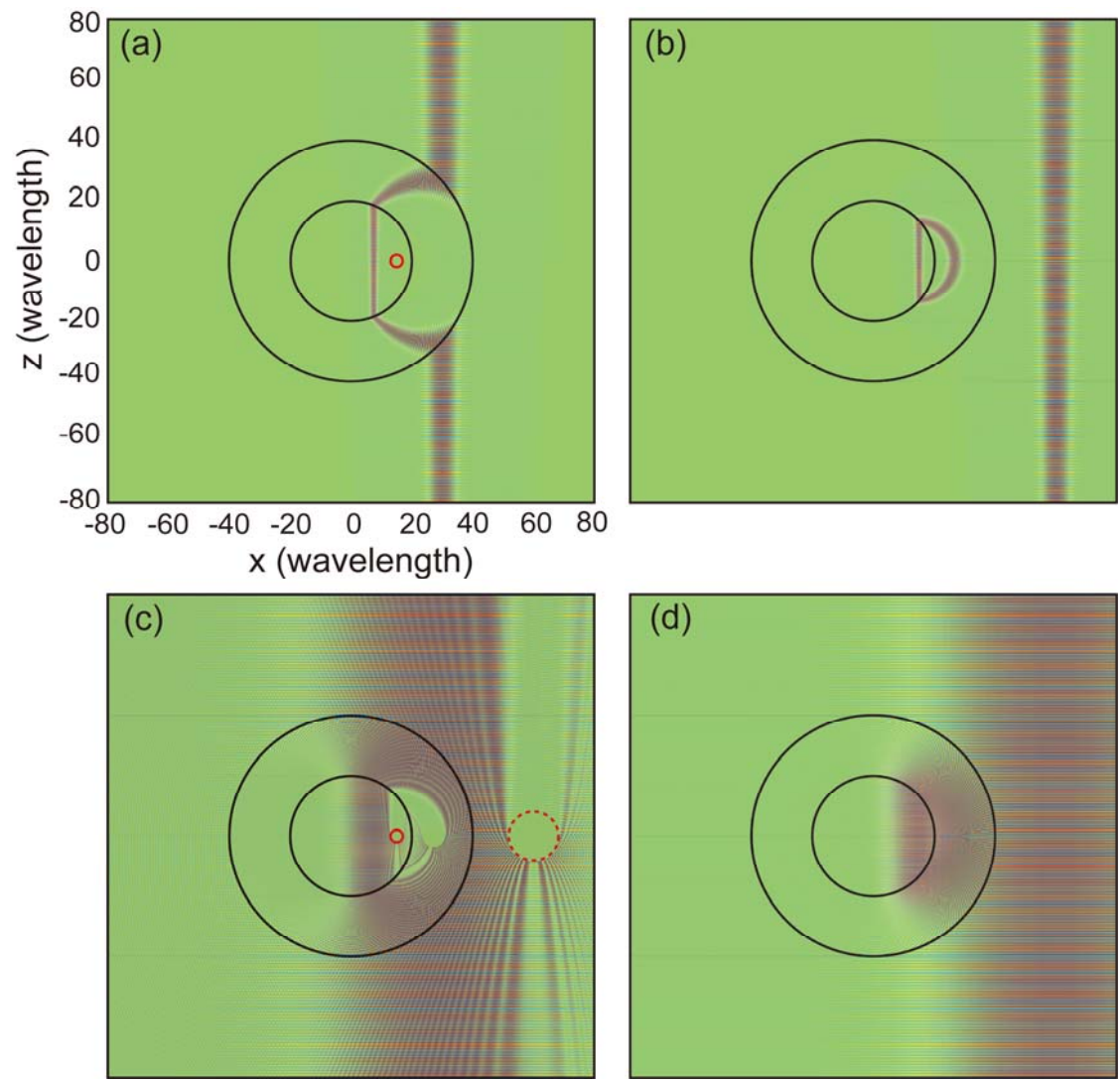